\documentclass[12pt,a4paper]{article}
\usepackage[latin1]{inputenc}
\usepackage{amsmath}
\usepackage{amsfonts}
\usepackage{amssymb}
\usepackage[pdftex]{graphicx}
\usepackage[left=2cm,right=2cm,top=2cm,bottom=2cm]{geometry}
\usepackage{setspace}
\usepackage{caption}
\usepackage{subcaption}

\author{D. A. Quint$^{1,2}$, A. Gopinathan$^{2*}$ \\and G. M. Grason$^{3*}$\\
Department of Bioengineering, Stanford University$^1$\\
Department of Physics, University of California, Merced$^2$ \\
and Department of Polymer Science and Engineering, \\
University of Massachusetts, Amherst$^3$\\
$*$ agopinathan@ucmerced.edu, grason@mail.pse.umass.edu}
\title{Shape selection of surface-bound helical filaments:  biopolymers on curved membranes}
\date{}
\begin{document}
\maketitle

\begin{abstract}
Motivated to understand the behavior of biological filaments interacting with membranes of various types, we study a theoretical model for the shape and thermodynamics of intrinsically-helical filaments bound to curved membranes.  We show filament-surface interactions lead to a host of non-uniform shape equilibria, in which filaments progressively unwind from their native twist with increasing surface interaction and surface curvature, ultimately adopting uniform-contact curved shapes.  The latter effect is due to non-linear coupling between elastic twist and bending of filaments on anisotropically-curved surfaces, such as the cylindrical surfaces considered here.  Via a combination of numerical solutions and asymptotic analysis of shape equilibria we show that filament conformations are critically sensitive to the surface curvature in both the strong- and weak-binding limits.  These results suggest that local structure of membrane-bound chiral filaments is generically sensitive to the curvature-radius of the surface to which it is bound, even when that radius is much larger than the filament intrinsic pitch. Typical values of elastic parameters and interaction energies for several prokaryotic and eukaryotic filaments indicate that biopolymers are inherently very sensitive to the coupling between twist, interactions and geometry and that this could be exploited for regulation of a variety of processes such as the targeted exertion of forces, signaling and self-assembly in response to geometric cues including the local mean and Gaussian curvatures.
\end{abstract}

\doublespacing
\section{Introduction}
 All living cells have a wide variety of filamentous biopolymers associated with the cell or nuclear membranes that play vital roles in biological functions specifically through their interactions with these membranes. In eukaryotes, for example, the actin cortex that resides just inside the cell membrane and is linked to it via a number of actin binding proteins, provides the cell with structural integrity and mediates signal transduction as well as cell-adhesion  \cite{cowin}. Other examples of membrane associated filamentous networks that provide mechanical stability include the actin-spectrin network in red blood cells and the nuclear lamin networks which are anchored to the cell and nuclear membranes by a number of specific binding proteins \cite{janmey, lamin}. Membrane-associated filaments can also be dynamic and exert forces such as the actin contractile ring in eukaryotes that provide the forces necessary for cell division \cite{cont}. Cell wall associated microtubules in plants play a significant role in guiding synthesis activity during the elongation phase of the cell cycle \cite{lloyd}. In bacteria, a similar role is accomplished by MreB which directs cell wall synthesis \cite{mreb,arkin,mrebdir} while the bacterial tubulin homolog FtsZ forms filaments that associate with the cell wall and function as contractile rings during division \cite{ftsz,arkin}. In many of these cases, the conformations and orientations of the membrane-bound filaments are critical for function as they exert forces and guide growth. \par
 Three very important physical parameters control these conformations and orientations: the helicity, or intrinsic twisted geometry, of the filaments; the strength of the binding interactions with the surface; and the local geometry or curvature of the surface. While conformations of polymers in contact with interfaces and surfaces have been well studied in the past \cite{polyint,polyint2}, the interplay of helicity and surface curvature introduce rich new behaviors. Freely-associating chiral polymers by themselves show novel phases in aggregates \cite{greg1,greg2,greg3,greg4} and we have shown recently that the interactions of a chiral polymer even with a flat surface dramatically restructures the filament shape and with non-trivial binding thermodynamics related to the Frenkel-Kontorova transition of incommensurate solids\cite{dqgreg}. We showed that there exists a critical binding strength, proportional to the torsional modulus of the filament and the square of its intrinsic twist, above which the filament unwinds to a zero-twist, surface-bound state and below which elastic energy of the filament peels off regions of strongly bound filaments through the proliferation of weakly-bound ``twist domains" . For filaments with anisotropic bending stiffnesses, this transition is then coupled to a dramatic change in the effective persistence length with the twist walls functioning as floppy joints. Recently it has also been shown that the interplay between twist elasticity and surface interactions  can lead to non-trivial, metastable 3D morphologies including loops and helices that lift off the surface \cite{sun_sm}. Thus, the conformations of surface bound helical polymers can depend sensitively on the binding interactions and this has implications not only for biopolymers {\it in vivo} but also for experimental studies of biopolymers immobilized on surfaces \cite{janmey} and for protein-based templated assemblies for nanotechnology applications \cite{santoso,peptoid}.  For example, amyloid fibrils which are essentially undesirable aggregates {\it in vivo} are responsible for a number of pathological conditions \cite{amyloiddisease} have been shown to be susceptible to membrane binding induced morphological changes in their twist states \cite{amyloid_fret} which has implications for the cytotoxicity {\it in vivo} but the coupling between binding and conformation via the twist may also be exploited for the design of amyloid based functional nano materials. \par
 While the coupling between chirality and binding produces a rich behavior even on flat surfaces, curvature is an essential feature of many of the surfaces of relevance {\it in vivo} and also potentially a desirable feature for surfaces used in a variety of  biotechnology applications. In this paper, we take the first steps towards understanding the combined effect of surface curvature, chirality and binding interactions on filament conformations.  Specifically, we aim to understand to the how the 3D equilibrium shape of filaments (curvature and torsion) is controlled not only by the strength of surface interactions, but also by the shape of the interface itself to which it is bound.  In section 2, we present a general model of a helical filament adsorbed to an anisotropically-curved (cylindrical) surface  and construct the shape Hamiltonian of the filament. In section 3, we present our numerical solutions of the the shape equations of motion for filaments on surfaces of variable binding strength and surface curvature.  Significantly, we showed that, on anistropically-curved surfaces, the equilibrium shape of filaments becomes increasingly curved as surface binding unwinds the helical twist of filaments, due to the non-linear geometrical interplay of twist and writhe for filaments on curved surfaces.  In section 4, we analyze the three key limits of the rich shape evolution of helical filaments on curved surfaces, beginning with the case of strong-surface binding and the transition from the strongly-bound, untwisted filament to the weakly-bound, twisted filament. We then look at the shape sensitivity of bound filaments in the limiting case of weak interactions with the curved surface, showing that even arbitrarily weak coupling between filament helicity and the surface lead to local changes of filament structure that are sensitive to surface curvature. In section 5 we discuss the implications of our results and potential experimental measurements.

\section{Model}
\begin{figure}[h]
\includegraphics[scale=0.44]{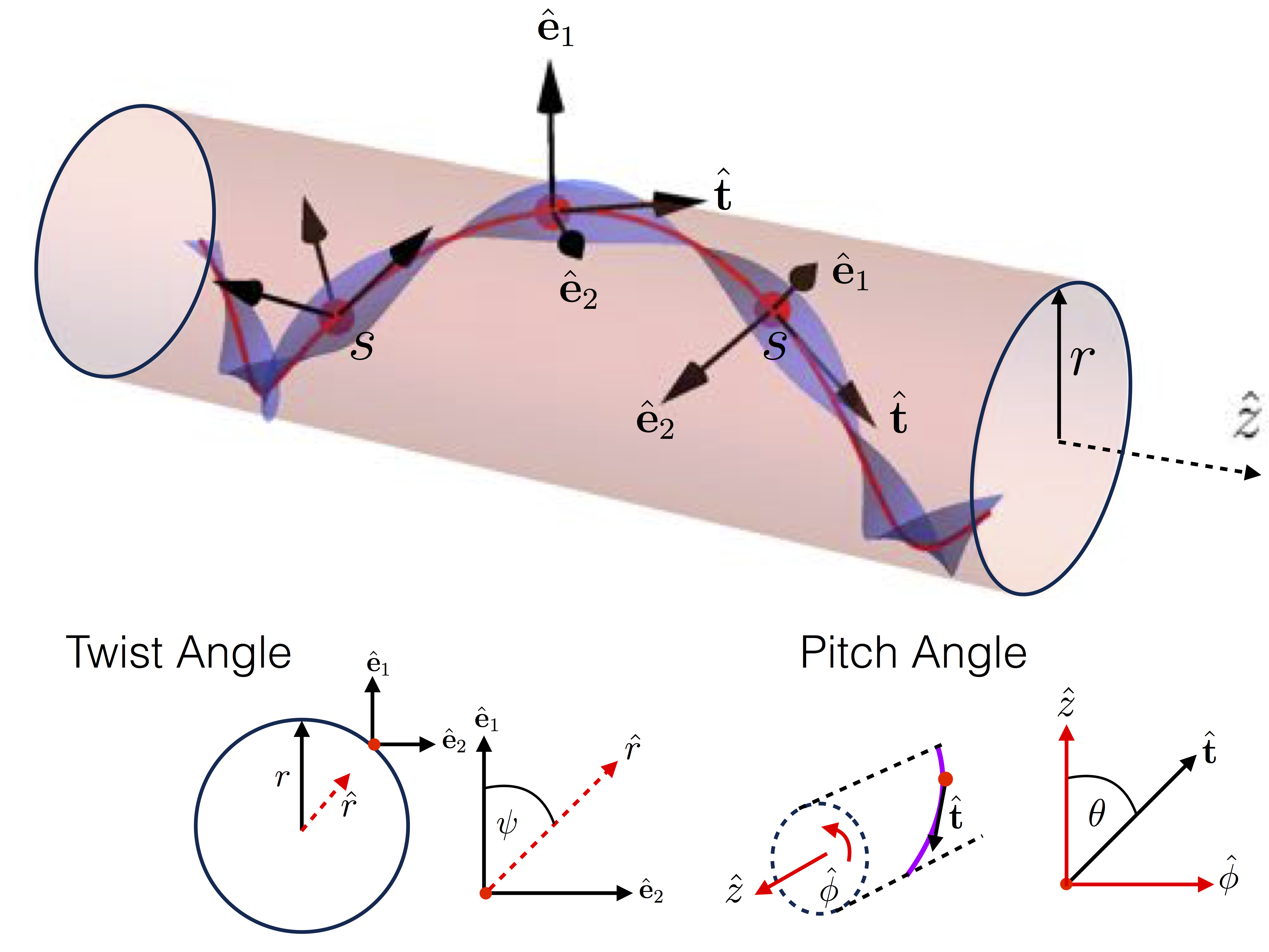}
\caption{A diagram depicting the model which we study. The triplet of axes are the filament tangent vector ($\hat{\mathbf{t}}$) which is related to the pitch angle of the helix and then the two material frame coordinates ($\hat{\mathbf{e}}_1, \hat{\mathbf{e}}_2$) which are related to the twist angle of the filament. The radius of the cylinder is $r$. }  \label{fig:fig1}
\end{figure}

Our model considers a thin filament of length $L$ that has a preferred intrinsic helical twist around its centroid of $\omega_0$ [deg/len.].  To depict the microscopic anisotropy of the filament, it is illustrated schematically as a helical ribbon in Fig.~\ref{fig:fig1}. We assume that filament backbone ${\bf r}(s)$ is localized to a cylindrical membrane of fixed radius $r$, the simplest model of an extrinsically-curved surface.   The local geometry of the bound filament is described by its tangent $\hat{\mathbf{t}}(s)$ (see figure \ref{fig:fig1}),
\begin{equation}
\partial_s \mathbf{r} = \hat{\mathbf{t}} = \cos \theta \hat{z} +\sin \theta \hat{\phi}.
\end{equation}
Here $\theta \equiv \theta(s)$ is the ``pitch" angle between the filament and the long axis of the cylinder, $s$ is the position along the filament backbone, and $\hat{z}$ and $\hat{\phi}$ describe the local longitudinal and azimuthal directions on the cylinder. To describe the twist degree of freedom, we choose two orthonormal unit vectors defining the material frame ,
\begin{equation}
\hat{\mathbf{e}}_1 = \cos\psi \hat{r} + \sin\psi(\hat{\mathbf{t}}\times \hat{r}),
\end{equation}
and
\begin{equation}
\hat{\mathbf{e}}_2 = -\sin\psi \hat{r} + \cos\psi(\hat{\mathbf{t}}\times \hat{r}).
\end{equation}

The angle $\psi \equiv \psi(s)$ is the angle between one ``face" of the helical filament (for example, the wide face of the ribbon) and the local normal to the cylinder, $\hat{r} = \hat{\phi} \times \hat{z}$. Using these three coordinate definitions we can compute the curvature of the filament in both of the principle material directions, $\kappa_i = \hat{\mathbf{e}}_i\cdot \partial_s \hat{\mathbf{t}}$, as well as the rate of twist of the material frame,  $\omega = \hat{\mathbf{e}}_1 \cdot \partial_s \hat{\mathbf{e}}_2$ (i.e. the model assumes a ``left-handed" convention for filament twist). The total curvature is given as,  [details in the supplementary material]

\begin{equation}
\kappa^2 = (\theta^\prime)^2 + \frac{\sin^4\theta}{r^2}
\label{eq:curve}
\end{equation}

where the first and second contributions derive from the geodesic and normal curvatures of the filament on the cylinder.  Similarly, the filament twist decomposes into two contributions,
\begin{equation}
\omega = \psi^\prime - \frac{\sin(2\theta)}{2r},
\label{eq:twist}
\end{equation}
where the first term represents the rotation of the material frame with respect to the local surface normal, while the second term derives from the rotation of the surface normal along the filament tangent.  

The elastic mechanical energy stored in the filament
\begin{equation}
E_{mech} = \frac{1}{2}\int_0^L\hspace*{2pt} ds \Big[ C\kappa^2 +K (\omega-\omega_0)^2 \Big],
\label{eq:mechE}
\end{equation}
arises from two sources, when it is bound to the cylinder; (1) bending energy, which is proportional to the local curvature (eqn. \ref{eq:curve}) the magnitude of which is set by the elastic constant $C$; and (2) the torsional elastic cost for deviations from the intrinsic twist of the filament $\omega_0$, which we assume is harmonic and whose magnitude is set by $K$, the elastic twist stiffness \cite{dqgreg}.  In the absence of external forces  on the filament, the shape of the filament is straight ($\theta=0$) with a twist rate of $\omega_0$. Here, we do not consider spontaneous curvature in this treatment and neglect anisotropy in bending moduli for $\hat{\mathbf{e}}_1$ and $\hat{\mathbf{e}}_2$ directions, though the model can easily be extend to consider these effects.
To associate the cylindrical wall with the filament, we consider that the filament possesses strong binding domains distributed along its contour that are in register with the intrinsic twist of the filament.  For the schematic example, ``strong binding" may be considered along the wider adhesive face shown in fig.~\ref{fig:fig1}. Furthermore, we envision that it is these domains (faces with $\psi = n \pi$) that favor contact with the cylindrical wall at all times as a opposed to the ``off-face" binding (i.e. $\psi = (n+1/2)\pi$).  As a minimal model of the interaction of the helical symmetry of the filament with the membrane, we introduce a periodic potential with a strength $V$,
\begin{equation}
E_{bind} = \frac{V}{2} \int_0^L \hspace*{2pt}ds \sin^2\psi.
\label{eq:bindE}
\end{equation}
Note the $\psi \to \psi + \pi$ symmetry is consistent with local $C_2$ symmetry of a double-helical filament cross-section, though it is straightforward, in principle, to modify the interaction according to any $n$-state helical symmetry.  

Combining the mechanical and binding energies along with eqns. \ref{eq:curve} and \ref{eq:twist} we arrive at the Hamiltonian that describes the energetics of the conformational phase space that the filament can sample from when it is bound to the cylinder. 

\begin{equation}
\mathcal{H} = \frac{1}{2} \int_0^L \hspace*{2pt}ds \Big[ C(\theta^\prime)^2 + C\frac{\sin^4\theta}{r^2} +K\bigg(\psi^\prime -\frac{\sin(2\theta)}{2r}-\omega_0\bigg)^2 +V\sin^2\psi \Big]
\label{eq:hamil}
\end{equation}

By inspection we see that helical filaments are frustrated by surface binding.   On one hand, it is not possible even in the undeformed case ($\theta=0,\psi^\prime =\omega_0$) for the helical filament to maintain ideal contact with the cylinder surface since the point of contact between the cylinder and the filament binding domain will happen periodically at a distance of $\omega_0/\pi$.   On the other hand, maintaining uniform, ideal contact ($\psi = n \pi$) leads generically to an elastic cost due to the preferred intrinsic twist.  However, unlike the case of planar substrates ($r \to \infty$) studied previously, any helical tilt of filaments on curved surfaces can relax the frustration through bending (i.e. $\theta \neq 0$).  This is due the geometric rotation of the surface normal along tilted paths $\theta \neq n \pi/2$.  Hence, while the transition from weakly-bound twisted filaments to strongly-bound untwisted filaments on flat surfaces is described by mathematics identical to the Frenkel-Kontorowa transition, on curved surfaces, the filament tilt acts a gauge field coupled to the twist degree of freedom.  As we find below, the strength of the coupling of the filament tilt to the twist, and its effect on the binding thermodynamics, is controlled by the dimensionless surface curvature, $(\omega_0 r)^{-1}$.

\section{Shape transitions:  numerical solutions}

In this section we analyze the equilibrium shapes of bound helical filaments conformations for surface binding potential $V$, for surfaces of varying surface curvature, $(\omega_0 r)^{-1}$, and filaments of varying ratios of bend to twist stiffness $C/K$. For a given set of set parameters, $V$, $\omega_0$, $r$, $C$ and $K$, we consider equilibrium shapes for filaments of arbitrary (unlimited) length on infinite-length cylinders.  Equilibrium shapes satisfy the following equations of motion, corresponding respectively to torque balance about the surface normal, 
\begin{equation}
C\theta'' =2C\frac{\sin^3\theta \cos \theta}{r^2} -K\frac{\cos(2\theta) }{2r}\Big(\psi^\prime -\frac{\sin(2\theta)}{2r}-\omega_0\Big) ,
\label{eq:tilt0}
\end{equation}
and about the filament tangent,
\begin{equation}
K\Big(\psi^\prime -\frac{\sin(2\theta)}{2r}\Big)^\prime=\frac{V}{2}\sin(2\psi) .
\label{eq:twist0}
\end{equation}
For the case of infinite-length filaments, we search for solutions that are periodic over an arc-distance $2{\cal L}$, and optimize the solutions with respect the total energy per unit length.   Because filament tilt compensates for frame rotation when $\psi'(s)<\omega_0$, we assume that for equilibrium shapes $\psi(s)$ and $\theta(s)$ solutions remain ``in phase" such that the magnitude of filament tilt reaches a maximum (minimum) at positions where the rotation rate $\psi'(s)$ is at a respective minimum (maximum).  Hence, we solve eqs.~\ref{eq:tilt0} and  \ref{eq:twist0} numerically, subject to the following boundary conditions,
\begin{equation}
\psi(0) = 0; \psi({\cal L}) = \pi; \theta'(0)=\theta'({\cal L})=0 .
\end{equation}
In practice, these equations are solved for $\psi'(0)\equiv \psi'_0$ is fixed and eqs.~\ref{eq:tilt0} and  \ref{eq:twist0} are solved via a standard shooting method to determine the value of initial tilt, $\theta(0)$, and half-period ${\cal L}$.  The energy per unit length of solutions is calculated via eq.~\ref{eq:hamil} and minimized with respect to $\psi'_0$, which is equivalent to minimization over ${\cal L}$.

\begin{figure}[h]
\begin{center}
\includegraphics[scale=0.44]{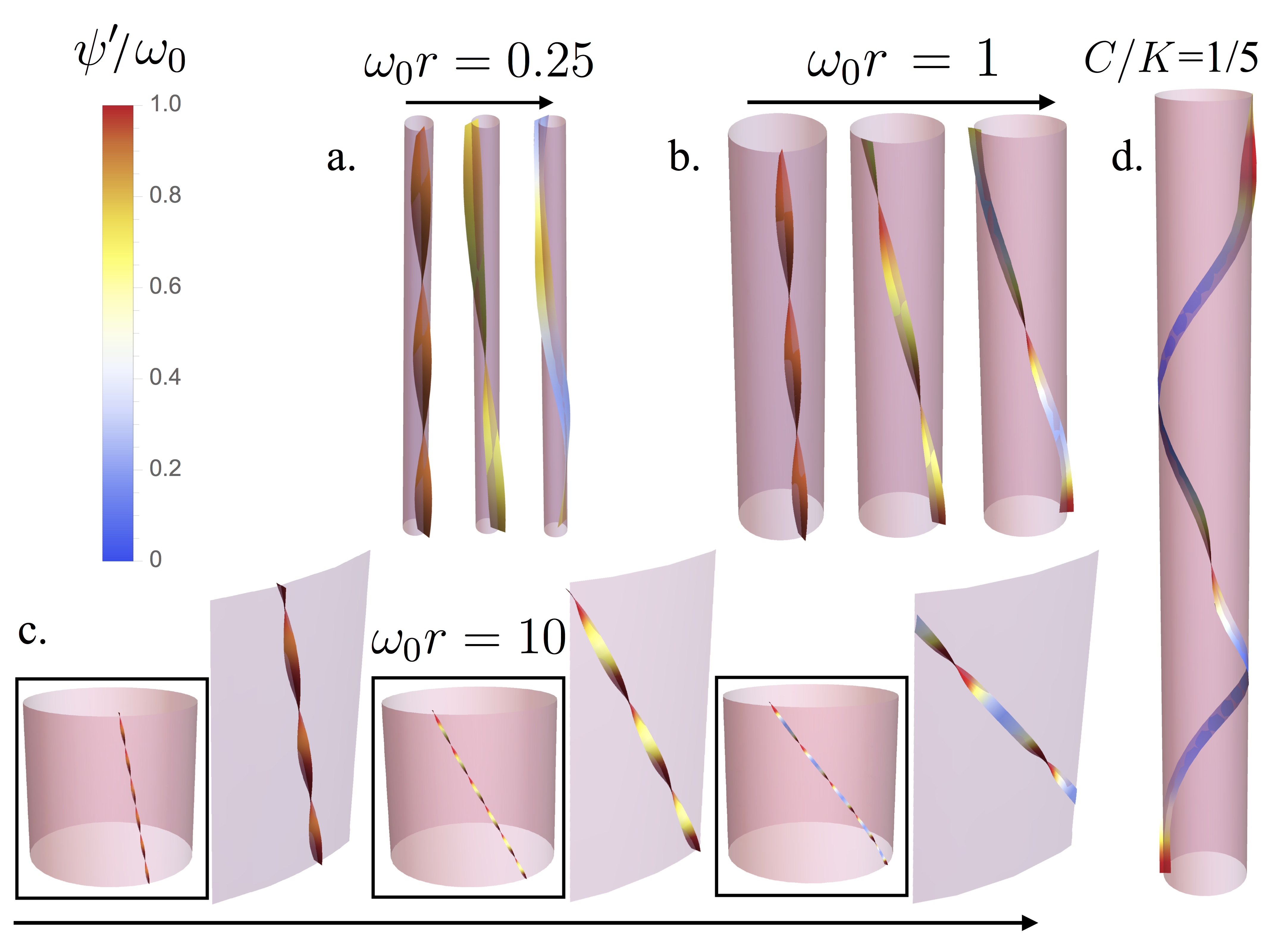}
\caption{The numerically obtained shape solutions for filaments with isotropic elastic properties $C=K$ (a-c) and d $C=K/5$.  For (a-c) from left to right, the three cases show increasing binding strength $V/V_c= 0.1, 0.5 \& 0.95$ (indicated by the black arrows), where $V_c$ is the critical binding strength for each $r$.  In (d) the binding strength is close to the critical value $V=0.95~V_c$, and exhibits pronounced axial straightening at twist domains. }\label{fig:fig2}
\end{center}
\end{figure}

In Fig.~\ref{fig:fig2} we present graphical portraits of the ``phase diagram" to summarize the variation of equilibrium filament shape in the parameter space spanned by surface interactions and surface curvature. In particular we show shape solutions for filaments with isotropic elastic properties $C=K$ for three values of reduced curvature:  $\omega_0 r =10$ (low curvature), $\omega_0 r =1$ (intermediate curvature) and $\omega_0 r =0.25$ (low curvature). For each case we see that as the interactions with the surface get stronger the filament that originally prefers an axial orientation with native twist begins to untwist and the coupling with curvature causes the pitch angle to rise. Ultimately, above some critical binding strength $V_c$, the filament is fully unwound assumes a constant ``face-on" configuration with uniform tilt angle $\theta$.

\begin{figure}[h]
\begin{center}
\includegraphics[scale=0.44]{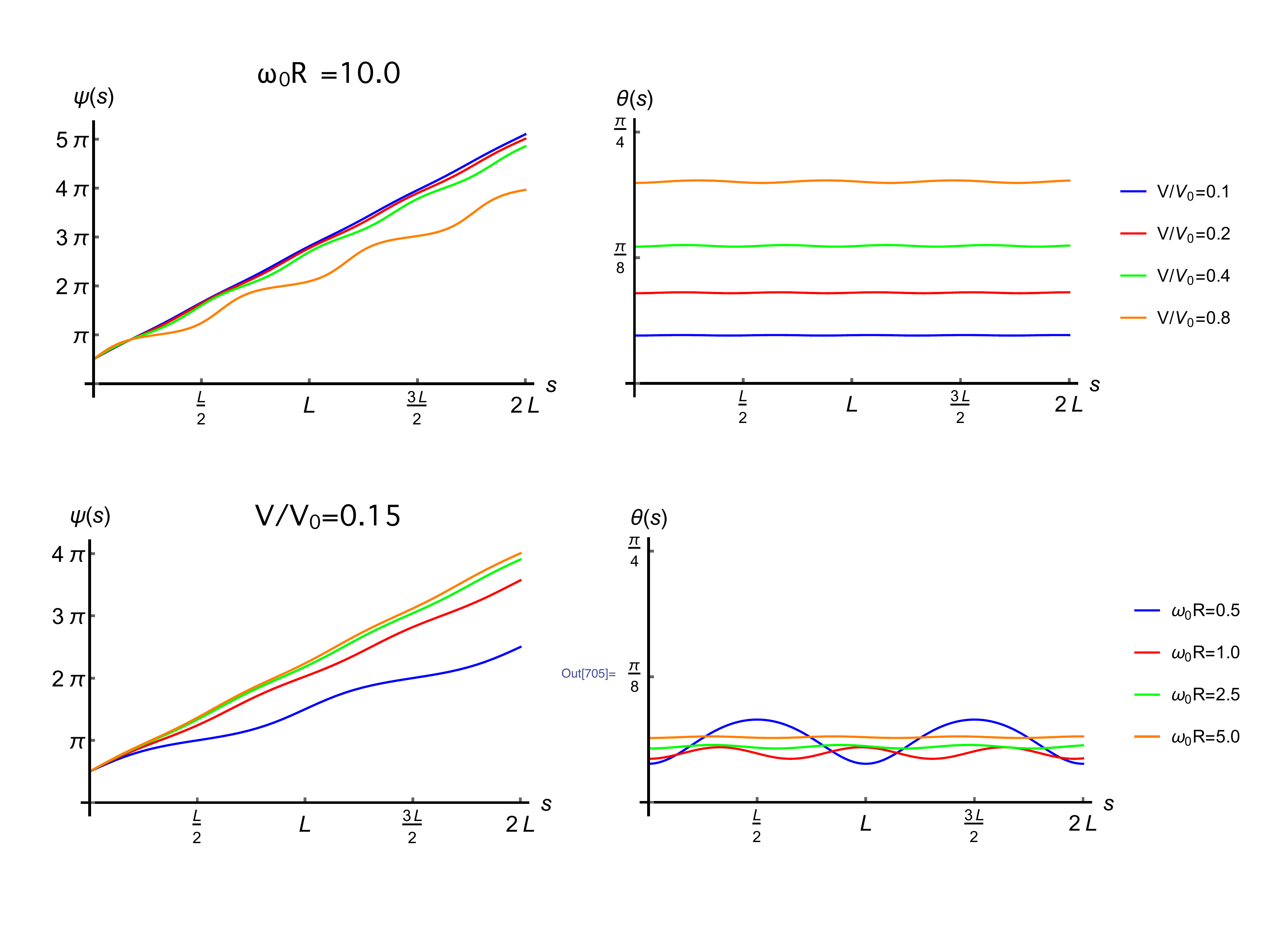}
\caption{Plots of filament twist $\psi$ and tilt $\theta$ for fixed curvature $\omega_0 r$ and varying values of the potential $V$ (top) and for fixed potential and varying curvature (bottom)   }\label{fig:fig3}
\end{center}
\end{figure}

To probe these trends more quantitatively we show, in fig.~\ref{fig:fig3}, the profiles of $\psi(s)$ and $\theta(s)$ for four values of the surface potential for a fixed value of curvature $\omega_0 r =10$.  We see that the progression of rotation and tilt angle profiles with $V$ shows a qualitatively similar sequence to what we observed from Fig.~\ref{fig:fig2}.  As $V \to 0$, the solutions approaches the elastically favorable intrinsic twist, $\psi(s) \to \omega_0$, and straight backbone, $\theta(s) \to 0$.  As $V$ increases, surface binding slows (speeds) the rate of $\psi$ rotation near the minima (maxima) of the surface contact, leading to an oscillatory profile of with a somewhat decreased mean value of $\langle \psi'(s) \rangle < \omega_0$, simultaneous with an increase in the magnitude of the mean tilt angle $\langle \theta(s) \rangle$.  As values of the surface potential approach a critical value ($V_c\simeq V_0$ the solutions rapidly evolve towards a critical unwinding transition via inhomogeneous structures characterized by rapid jumps in $\psi(s)$ by $\pi$. These correspond to ``twist domains", separated by increasing stretches of strongly binding and nearly constant $\psi \approx n \pi$.  Again, this progression towards the critical $V$ is accompanied by further increase in $\langle \theta(s) \rangle$, with the higher degree of tilt on the larger-radius surface.  

Beyond this critical value of surface binding, filaments adopt an uniformly unwound, helical conformation with $\psi(s) = 0$ and $\theta(s) = {\rm const.}$, with shape independent of surface potential in this large-$V$ region.  In fig.~\ref{fig:fig3}, we also show how $\psi(s)$ and $\theta(s)$ profiles vary when $V$ is held fixed and the curvature $\omega_0 r$ is varied. Again we see that higher curvatures more strongly influence both the pitch and twist of the filaments.

\begin{figure}[h]
\begin{center}
\includegraphics[scale=0.44]{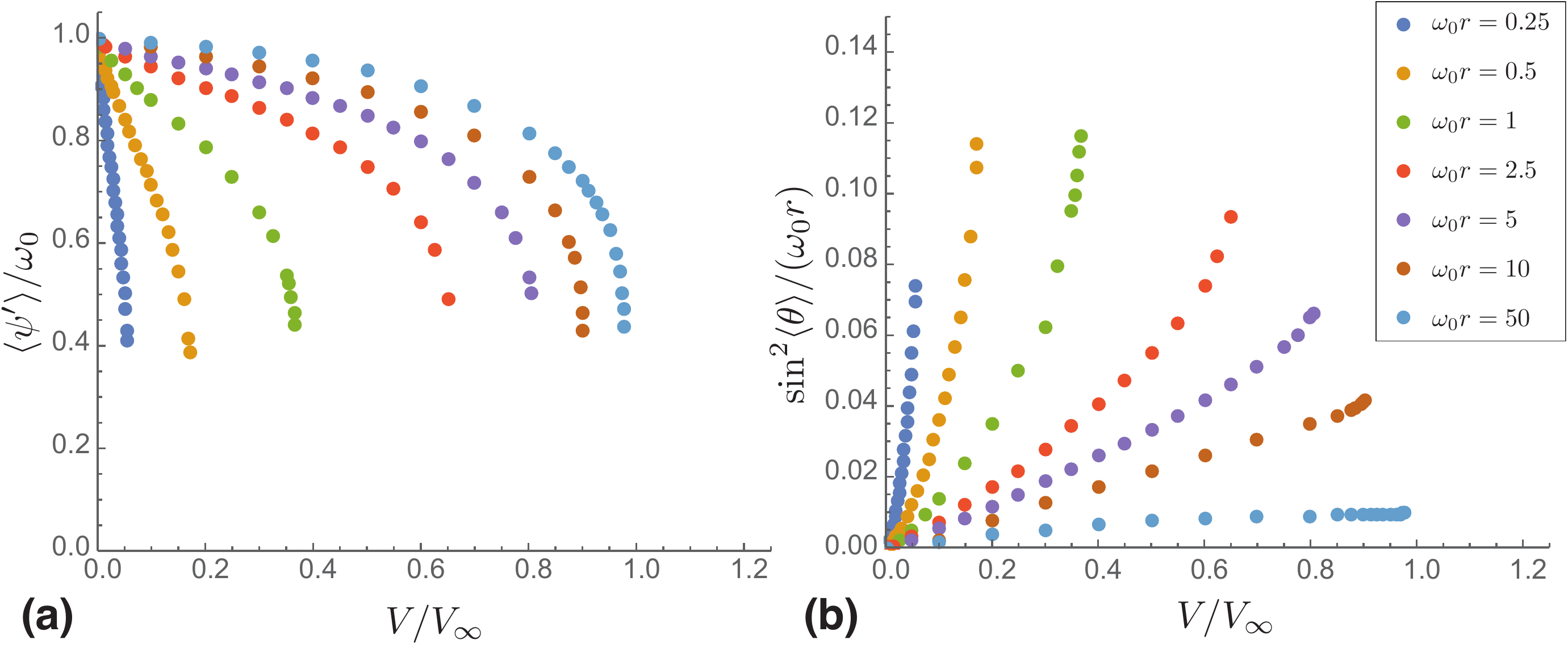}
\caption{Plots of renormalized twist rate $\langle \psi' \rangle$ and tilt $\sin^2\langle \theta \rangle/r$  for equal bending and torsional stiffness $C=K$}\label{fig:fig4}
\end{center}
\end{figure}

In Fig. ~\ref{fig:fig4} we plot the mean rotation rate $\langle \psi' \rangle$ and   $\sin^2\langle \theta \rangle/r$ (a measure of the net curvature of the filament) as functions of surface-binding potential for a range of surface radii, $(\omega_0 r)=0.25-50$.  Again, these each show a gradual decrease in the net rotation of the filament for small $V$ followed by a precipitous drop to $\langle \psi' \rangle=0$ at a critical surface potential $V_c$ which decreases with increasing surface curvature.  For the smallest curvature studied,  we find that $V_c \simeq  \pi^2 K \omega_0^2/4$, approaching the asymptotic limit of binding on flat surfaces.  Compared to largest curvature ($\omega_0 r =0.25$) we observe $V_c \simeq 0.05 (\pi^2 K \omega_0^2/4)$, a dramatic reduction in threshold surface interaction needed to unwind the filament.     

\begin{figure}[h]
\begin{center}
\includegraphics[scale=0.44]{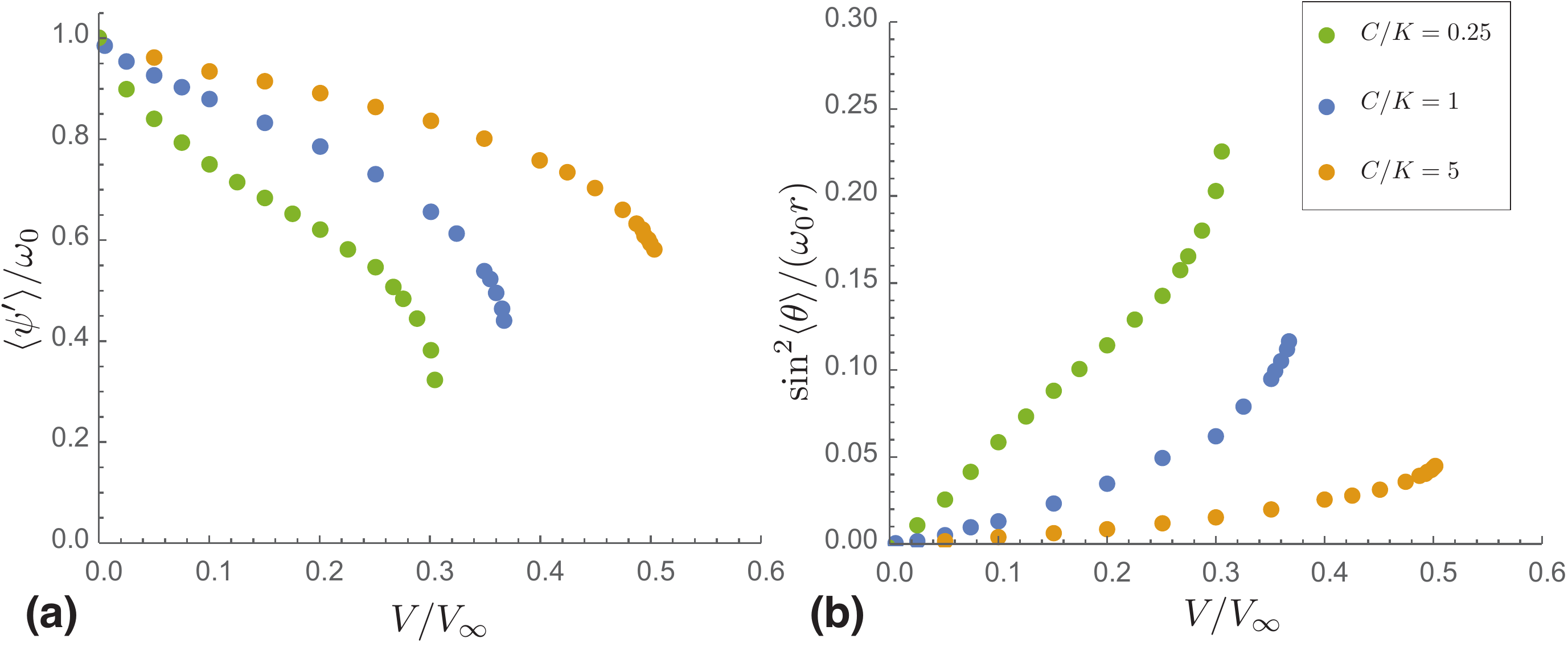}
\caption{Plots of renormalized twist rate $\langle \psi' \rangle$ and  tilt $\sin^2\langle \theta \rangle/r$ for varying ratios of bending and torsional stiffness $C/K$ }\label{fig:fig5}
\end{center}
\end{figure}

We analyze further how the coupling between twist and tilt degrees of filaments varies not only with surface curvature, but with the ratio of bend to twist elastic constants, $C/K$.  In Fig. ~\ref{fig:fig5}  we plot $\langle \psi' \rangle$ and $\sin^2\langle \theta \rangle/r$ vs. $V$ at fixed curvature $\omega_0 r=1$, for three different elastic anistropies: $C/K=0.25, 1$ and $5$.  Due to the diminished effect of screening of the elastic cost of twist from helical bending (or tilt), increasing bending stiffness relative to twist stiffness shows an increasing in the threshold surface binding for unwrapping the filament.  In contrast, the weakening the bending stiffness relative to twist shows an increased sensitivity to surface binding and smaller $V_c$ (relative to $\pi^2 K \omega_0^2/4$).  

Overall, these results illustrate increasing the surface curvature and decreasing the relative bending stiffness $C/K$ lead to a marked increase in the susceptibility of filament shape (both twist and curvature) to surface interactions that couple to the helical symmetry of the filament.  

\section{Shape sensitivity to surface curvature}

In this section, we describe the mechanisms of shape-selection for surface bound helical filaments, and in particular, the sensitivity of filament shape to surface curvature.  We begin by analyzing the limiting case of strong-surface binding and the transition from the strongly-bound, unwound filament to the weakly-bound, twisted filament.  We follow with an analysis of the shape sensitivity of bound filaments in the limiting case of weak interactions with the curved surface.

\subsection{Strong-binding and unbinding transition}

\begin{figure}[h]
\begin{center}
\includegraphics[scale=0.44]{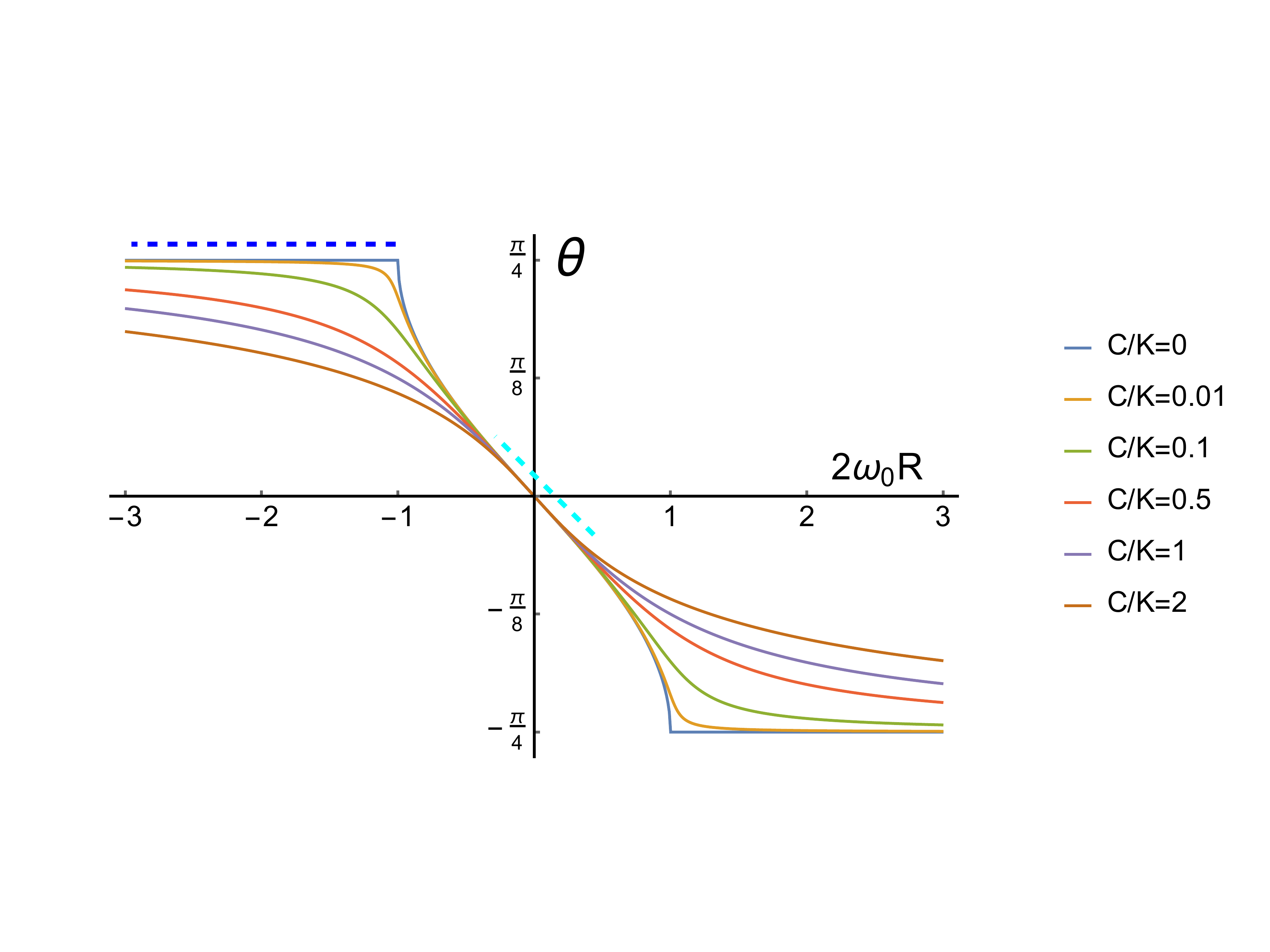}
\caption{Plots of the equilibrium tilt $\theta_0$ as a function of curvature $2\omega_0 r$ in the strong binding limit for varying ratios of bending and torsional stiffness $C/K$ }\label{fig:fig6}
\end{center}
\end{figure}

We begin by considering the shape of filaments in the $V \to \infty$ limit, where the strong-binding face or interactions sites on the filament are not free to peel way from the surface.   Assuming that filament orientational locally optimizes surface cohesion ($\psi=0$), equilibrium configurations correspond to helices of constant $\theta=\theta_0$, determined by the strong-binding equation of state,
\begin{equation}
2C\frac{\sin^3\theta_0 \cos \theta_0}{r^2} =K\frac{\cos(2\theta_0) }{2r}\Big(\frac{\sin(2\theta_0)}{2r}+\omega_0\Big)
\end{equation}
The tilt-equilibria are plotted in Fig.~\ref{fig:fig6} as a function of reduced twist, $2\omega_0 r$, showing a generic rise for $\theta =0$ to the tilt that provides the maximum rotation of the surface normal, $\theta=\pi/4$, and hence the largest possible relaxation of the twist elastic energy of the bound filament.  It is straightforward to show that tilt equilibria have the following asymptotic limits, 
\begin{equation}
\theta_0 = \left\{ 
	\begin{array}{ll}
		- \sin^{-1}(2\omega_0r)/2\simeq \omega_0 r &  2\omega_0 r \ll 1 \\
		-\pi/4 + \frac{C/K}{2 \omega_0 r-1+C/K} &  2 \omega_0 r \gg 1
	\end{array} \right.
	\label{eq:theta}
\end{equation}      
increasing linearly for small reduced twist, and saturating at $\theta_0=\pi/4$ for large twist.    We plot further in Fig.~\ref{fig:fig6}, the dependence of $\theta_0$ on the ratio of bend to twist elastic constants, $C/K$, which illustrates that the shape of strongly-bound filaments is determined not only by the degree of torsional strain, or  $\omega_0 r$, but also the relative cost of relaxing that strain through bending deformations.  

These constant-helix solutions have been previously studied as models of chiral filaments uniformly bound to cylindrical membranes, such as the bacterial cell well \cite{arkin}.  Here, we show the non-linear evolution of shape-equilibrium with increasing the strength of filament-surface interactions, where sufficiently weak interactions allow filament conformations peel away from uniform surface contact.  In a previous study, we addressed the transition from weakly-bound to strongly-bound helical filaments for planar interfaces.  Taking the limit of $r\to \infty$ in eq. \ref{eq:hamil} we arrive at the planar model, described by the energy
\begin{equation}
\lim_{r \to \infty} \mathcal{H}= \frac{1}{2} \int_0^L \hspace*{2pt} ds \bigg[ C |\theta'|^2 + K(\psi^\prime - \omega_0)^2 +V\sin^2(\psi)\bigg],
\label{eq:Hpsi}
\end{equation}
showing that bend and twist-orientation degrees of freedom decouple on flat surfaces.  As previously noted, the transition from strong- to weak-binding with decreasing surface potential $V$, maps on to the commensurate-incommensurate transition of the Frenkel-Kontorowa model of surface adsorption.  The ``unwound" filament with uniform $\psi=0$ (i.e. the commensurate state) is stable for $V>V_\infty$ where the critical potential is
\begin{equation}
V_\infty=\frac{\pi^2}{4}K \omega_0^2.
\end{equation}
Near to, but below this binding strength, localized jumps of $\psi$ by $\pi$ over a length scale proportional to $\sqrt{K/V}$ become stable in the equilibrium shape.  These localized ``twist walls" (``discommensurations" in the language of incommensurate solids) are separated by a characteristic distance ${\cal L}$ that diverges as the potential approaches its critical value from below ${\cal L} \sim - \ln(V_\infty-V)$.  As the binding strength is decreased far below $V_\infty$, distinct twist walls merge and the filament twist profile evolves continuously to the state of zero elastic strain and native twist ($\psi = \omega_0 s$) as $V \to 0$.  

\begin{figure}[h]
\begin{center}
\includegraphics[scale=0.75]{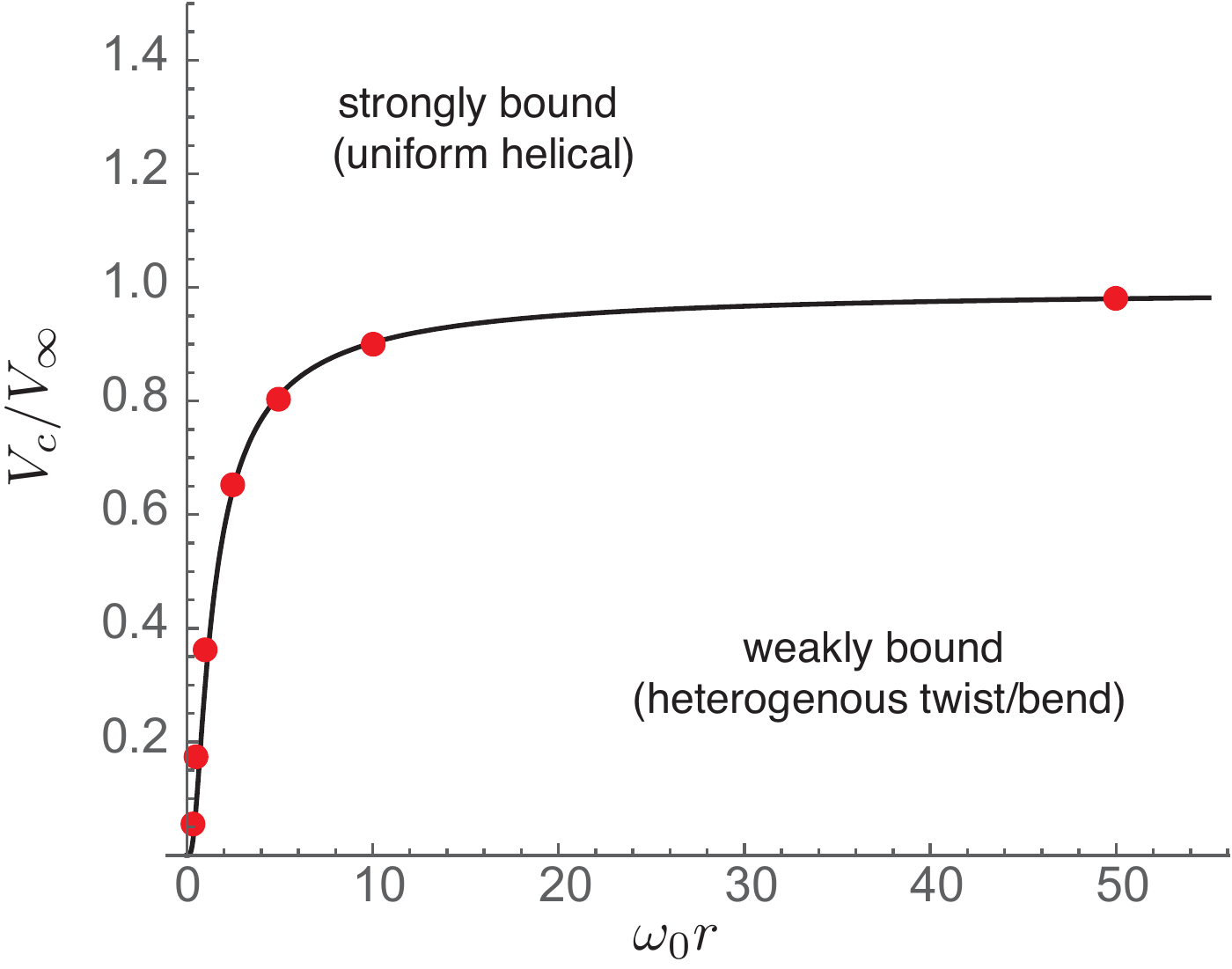}
\caption{Plot of critical potential $V_c/V_\infty$ as a function of curvature $\omega_0 r$.  The red points show the results of numerical solutions, while the solid curve is the prediction of ``renormalized" critical potential given by eq. (\ref{eq: renorm}). }\label{fig:fig7}
\end{center}
\end{figure}

Noting that the binding threshold separating strong-binding (uniform $\psi$, $\theta$) from weak-binding (non-uniform $\psi$, $\theta$) is found to be strongly dependent on the coupling of local filament orientation to surface curvature, we propose a simple generalization of the flat-interface analysis for the critical surface potential on curved surfaces.  When the uniform tilt solution of eq. \ref{eq:theta} is inserted into the elastic energy of eq. \ref{eq:hamil}, we note that the torsional strain is reduced from the native twist $\omega_0$ by $-\sin(2 \theta_0)/(2 r)$.  Therefore, the torsional loading of the filament on the cylinder is {\it reduced}  relative to a flat interface, and we expect the filament to be ``unwound" by a far weaker critical potential $V_c$, proportional to the square of a {\it renormalized twist} $\omega_{\rm eff} = \omega_0+\sin(2 \theta_0)/(2 r)< \omega_0$.  Hence, we estimate the dependence of the critical potential on dimensionless twist $\omega_0 r$,
\begin{equation}
\label{eq: renorm}
V_c(\omega_0 r) = V_\infty\Big(1+\frac{\sin (2 \theta_0)}{\omega_0 r}\Big)^2
\end{equation}
This estimate for the threshold binding strength is compared to numerical results for the transition between non-uniform and uniform $\psi$ solutions, show strong agreement from small to large curvature (fig.~\ref{fig:fig7}).  Using the solution of $\theta_0$ as $\omega_0 r \to 0$, we estimate that, for large curvature, the critical binding strength vanishes as, $\lim_{r \to 0} V_c \approx  V_\infty (\omega_0 r)^4$, due to the elimination of the elastic twist penalty to unwind the filament into perfect surface contact through small bending deformations.  In the opposite limit, where $\theta_0 \to \pi/4$ as $\omega_0 r \to \infty$ we find a continuous increase of the critical potential to planar threshold, $V_c \to V_\infty$.

\subsection{Curvature sensitivity for weakly-binding filaments}

In this section we illustrate the effect of surface interactions with helical symmetry of the filament for the limit of $V \ll V_c$, where the filament structure is only weakly perturbed by surface interactions with its helical symmetry.  The analysis is based on perturbative solution to filament shape equations for weak binding (see Appendix for details).  Here, we find the weakly perturbed filament conformation,
\begin{equation}
\psi(s) \simeq \omega s + \delta \psi (s); \theta(s) \simeq \theta_0 +\delta \theta (s) ,
\end{equation}
where $\omega$ and $\theta_0$ mean filament twist and tilt, respectively, while $\delta \psi (s)$ and $\delta \theta (s)$ represent (zero mean) longitudinal modulations of shape deriving from the position-dependent torques along the filament.  First consider this planar case $r \to \infty$ and vanishing coupling between twist and tilt.  In this case, expanding the equation of motion for small $\delta \psi$ the balance of torques about the filament axis sets $K \delta \psi'' \approx V \sin (2 \omega s)$.  This implies a modulation of filament twist in registry with the mean twist $\omega$ which decelerates (accelerates) the twist  where the filament-surface binding is locally minimal (maximal), with an amplitude proportional to $V/(K \omega^2)$.  Hence, the net energy gain per unit filament length due to this weak-correlation effect is roughly $-V^2/(K \omega^2)$, indicated a preference to unwind the natural pitch of the filament.  Balancing this preference is the elastic cost for altering the mean pitch of the filament $K(\omega-\omega_0)^2/2$, which for small $V$ gives the following parabolic dependence of torsional strain on potential.
\begin{equation}
\lim_{V \to 0} (\omega- \omega_0) =  -\frac{V^2}{16 K^2 \omega_0^3} \ {\rm for \ } r \to \infty .
\end{equation}
For filaments on curved cylindrical surfaces, we find two additional effects.  First, there is an additional contribution to the correlation energy, proportional to $-V^2/(C \omega^4 r^2)$ per unit length, due to enhanced ability of bound filaments to locally accelerate/decelerate the rotation of orientation though oscillatory ``wobbling" of the filament tilt to further optimize local surface contact.  The second, and perhaps more critical, difference is the screening of the elastic cost of mean torsional strain through tilt.  In contrast to the harmonic cost on planar surfaces, in the $V \to 0$ limit the elastic cost for deviations from native twist becomes much softer on curved surfaces, $C(\omega-\omega_0)^4r^2/2$ per unit length.  Optimizing for mean twist in $V\to 0$ limit, we find the torsional strain on curved surfaces
\begin{equation}
\lim_{V \to 0} (\omega- \omega_0)  =  - \bigg[\frac{V^2}{32 C \omega_0^3 r^2} \bigg( \frac{K^{-1}}{2} + \frac{3C^{-1}}{16 \omega^2 r^2}\bigg)\bigg]^{1/3}; \ {\rm for \  finite} \  r
\end{equation}
Therefore, we find the sensitivity to surface potential of helical filament shape on curved surfaces is critically different from flat surfaces.  

\begin{figure}[h]
\begin{center}
\includegraphics[scale=0.75]{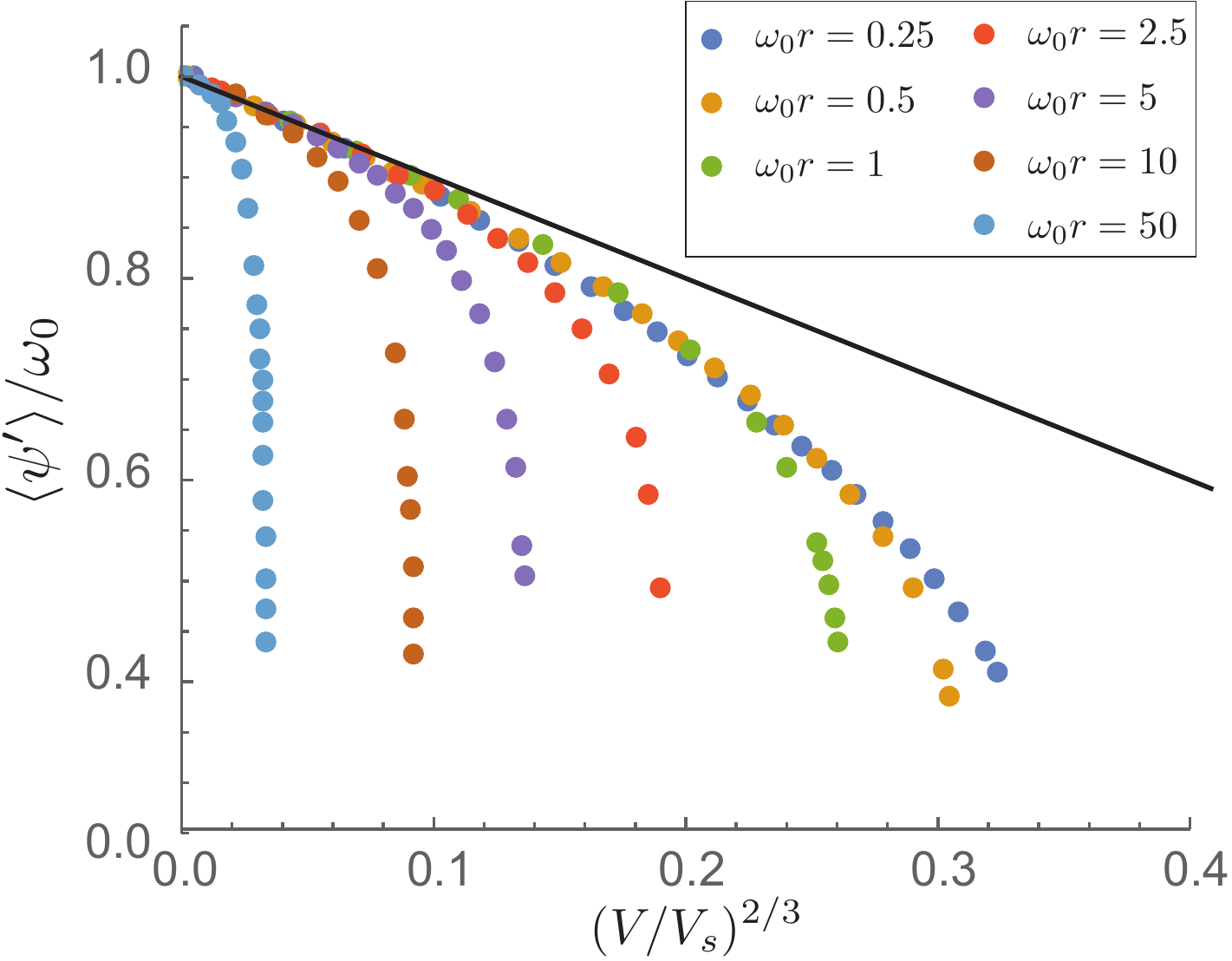}
\caption{Plot of torsional strain versus renormalized potential at small $V$. Straight black line indicates the $V^{2/3}$ scaling.  }\label{fig:fig8}
\end{center}
\end{figure}

Most notably, we find that the power-law dependence of torsional strain $\omega-\omega_0$ changes from the weaker $V^2$ on flat surfaces to  $V^{2/3}$ on curved surfaces.  Second, we observe that the torsional strain at small $V$ is strongly dependent on surface curvature $r$ as well as filament stiffness.  This highlights the remarkable fact that the local structure of the filament (its elastic strain) is generically sensitive to the {\it surface shape}, even in the asymptotic limit of weak surface interactions.  This accounts for the much more rapid decrease of filament twist observed for filaments as $\omega_0 r $ is decreased, shown in Fig.  ~\ref{fig:fig8}.  We verify the predicted dependence of torsional strain in the $V \to 0$ limit by replotting mean-twist $\langle \psi \rangle = \omega$ vs. the right-hand side of the equation above, showing universal agreement for $\omega_0 r$.  Finally, we note by comparing the respective strain predictions for flat curved interfaces, that we expect a crossover between the singular $V^{2/3}$ dependence at small potentials to the planar scaling $V^2$ at a characteristic binding strength, $V_s \approx (K^5 \omega_0^6/C r^2)^{1/4}$, indicating that the range of weak-binding where strain exhibits strong-curvature decreases with decreasing curvatures as $(\omega_0 r)^{-1/2}$.  

\section{Discussion}
Our results show that, for surface bound biopolymers, their elasticity and chirality can combine in non-trivial ways with surface interactions and surface geometry to determine equilibrium morphology. For twisted filaments interacting with a flat surface, we previously showed \cite{dqgreg} that there exists a critical interaction strength $V_\infty=\frac{\pi^2}{4}K \omega_0^2$ above which the filament exists in an untwisted, strongly bound configuration that maximizes the interaction energy at the expense of torsion. Here we show that the existence of surface curvature introduces an extra degree of freedom, whereby the filament can transition to the untwisted state at lower values of the interaction strength by essentially relieving torsional strain via writhing to maintain surface contact. This results in a lowered value for the critical interaction strength - $V_c(\omega_0 r) = V_\infty\Big(1+\frac{\sin (2 \theta_0)}{\omega_0 r}\Big)^2$ and a spontaneous preference for helical morphologies even in the absence of any spontaneous curvature simply due to the interplay between twist, writhe, surface interactions and geometry. \par
This is of particular interest in the bacterial context where cytoskeletal filaments like FtsZ and MreB have specific orientations relative to cylindrical cell bodies that prove essential in applying forces and templating growth in the correct locations. In order to assess the significance of our results for bacterial cytoskeletal filaments we consider the specific values of the elastic constants and geometric parameters that are involved and where this would situate these systems in our phase diagram \ref{fig:fig7}. For FtsZ, the intrinsic twist, as reported by recent MD simulations \cite{jen_pnas, tarazona_sm}, is in the range of $3^{\circ}-20^{\circ}$ which we take to be of the order of $\sim 10^{\circ}$ per monomer (of size $5$ nm) resulting in $\omega_0 \sim 0.03$~nm$^{-1}$. Given that cell radii for {\it {E.coli}} bacteria are typically 100-500 nm, we get $\omega_0 r \sim 3-15$. For single protofilaments of MreB, MD results \cite{alex_pnas} indicate a smaller angle of about  $3^{\circ}$ per monomer yielding $\omega_0 r \sim 1-5$.  To estimate the critical potential $V_\infty$, we additionally require the torsional modulus $K$. The variance, $\sigma^2$ of the fluctuations in the twist angle from independent MD simulations, gives a consistent value of $\sigma^2 \sim 15$ degree-squared for FtsZ \cite{jen_pnas, tarazona_sm}. From this one may estimate the torsional rigidity as $K=k_B T \ell /\sigma^2 $, where $ \ell \sim 5$nm, is the monomer size, giving us $K \sim 1200 k_B T$ nm. For single MreB protofilaments, a similar analysis gives $K \sim 2000 k_B T$nm \cite{alex_pnas}. This gives us a value for $V_\infty=\frac{\pi^2}{4}K \omega_0^2 \sim 2 k_BT/$nm for FtsZ and $0.5 k_BT$/nm for MreB. It is to be noted that these parameters could change for high order assemblies that occur {\it in vivo}.
The actual interaction potentials between FtsZ/MreB and the membrane are complex and mediated by multiple linkers. An estimate of the binding affinity of FtsZ for ZipA linker coated substrates \cite{ftsz_jbc} allows us to approximate the interaction strength at about $1 k_BT$/nm yielding a ratio $V/V_\infty \sim 1/2$. In addition to specific interactions from linking molecules, non-specific interactions such as electrostatics could also contribute significantly to the net interaction strength. Given that FtsZ (and also MreB) have linear charge densities comparable to actin ($\sim 4$e/nm) \cite{popp_jbc} and that reasonable membrane charge densities under physiological conditions could give rise to $V_{el} \sim 1-2 k_BT$/nm for actin \cite{dqgreg}, we anticipate that $V_{net}/V_\infty$ could range from 0.2 to 1.5 for all cases considered. It is to be noted that these ranges of values for $\omega_0r$ and $V/V_\infty$ cover the region corresponding to the knee of the curve in fig.\ref{fig:fig7}, indicating that these filaments are highly sensitive to changes in {\it the global surface geometry}, twist state as well as changes in the interaction potential, wherein small changes in these parameters (e.g. the radius of curvature, $r$) can cause dramatic changes in morphology and orientation. \par
Another interesting feature of this novel coupling emerges when we consider that filaments can have spontaneous curvature as well. A transition in the twist state can the be coupled to a reorientation or even the emergence of a plane of spontaneous curvature that can then lead to the exertion of forces \cite{tarazona_sm}. It has been observed with AFM experiments on FtsZ polymerized on mica \cite{hamon_langmuir} that they can exist in two states - as long curved filaments or short straight filaments, which could arise from different torsional states in the two populations. Experiments on FtsZ adsorbed to curved lipid surfaces \cite{arumugam_angchem} have also shown that the resulting orientations cannot be accounted for by spontaneous curvature and must include a coupling to twist. These experimental observations lend support to our model and also suggest that the non-linear interplay between geometry, twist and surface interactions can be used to regulate force production.\par
The coupling between filament bending and twist studied here for cylindrical surfaces, will arise on any surface which is anisotropically curved, a fact that has important biophysical implications.  For a strongly-bound filament whose material frame is locked to the local tangent frame of an binding surface, the twist has the form $2(\kappa_1-\kappa_2) \sin(2 \theta)$, where $\kappa_1$ and $\kappa_2$ are the principle curvatures of the surface and $\theta$ is the angle between the filament tangent and the first principle axis.  This formula shows that a strongly-bound filament can maintain maximal twist by aligning at $45^\circ$ with respect to the principle axes of an anistropically-curved surface.  Combining this with the normal curvature for a filament $\kappa_1 \cos^2 \theta + \kappa_2 \sin^2 \theta$, we see that a surface of {\it negative Gaussian curvature} where $\kappa_1 \kappa_2 \leq 0$ allow filaments to bind with particularly low elastic energy because there is always a straight path between the principles axes when principle curvatures have opposite sign.  For example, for minimal surfaces, $\kappa_1=-\kappa_2$, binding at $45^\circ$ with respect to principle axes achieves the maximal twist of a strongly bound filament ($\pm 2 \kappa_1$) while requiring {\it no bending} of the filament backbone.  These arguments suggest binding of helical filaments will be {\it generically favored in regions of negative Gaussian curvature}.  This mechanism may have implications for the localization of bacterial cytoskeletal filaments to cell-wall geometries with negative curvature, such as the localization of crescentin that occurs in crescent shaped {\it Caulobacter Crescentus} \cite{crescentin}, or that proposed for MreB in either plastically deformed rod-like bacterium \cite{ariel_pnas} or in maintaining rod shapes in growing cylindrical bacteria \cite{kc_gaussian}.\par
The fact that many biopolymers, both eukaryotic and prokaryotic, that are composed of proteins will have closely related properties indicates that this kind of coupling between twist, interactions and geometry could be exploited for regulation of a variety of processes. For example adsorbing to the surface and unwinding in could expose moieties on the monomer surfaces that could trigger biochemical pathways in response to geometric cues, such as the presence of regions with different curvatures. This could be of use in directing function to geometrically defined regions such as mid plane constrictions for FtsZ or regions where extra cell wall synthesis machinery mediated by MreB (as suggested in the preceding paragraph) could be directed.  The transition could also be regulated by changing  the intrinsic twist or even just the linear charge density by post translational modifications.  Another interesting possibility is that transition could be accompanied by the exposure of regions of the monomer surface that promote bundling or in-plane aggregation. This could then lead to an auto-catalytic accumulation of filaments in targeted regions. One could imagine that such a mechanism would be valuable in the design of self-assembly pathways. For example, amyloid fibrils have been found to untwist upon interactions with lipid membranes \cite{amyloid_fret, sun_sm}, which could in turn affect their aggregations and be exploited for for the design of amyloid or protein based functional nano materials \cite{santoso} on arbitrarily curved surfaces that could be responsive to changing geometries. Finally it is to be noted that many of these cytoskeletal filaments are in a state of dynamic turnover and it has been shown \cite{sun_bj} that coupling the kinetics of filament polymerization with cell wall growth and mechanics can lead to non-trivial regulatory mechanisms. It would be interesting to consider the role of the coupling of our untwisting transition to these mechanisms.\par

\section{Author Contributions}
AG and GG designed the study. DQ, AG and GG performed the research and wrote the paper.

\section{Acknowledgements}
DQ and AG were supported by a James S. McDonnell Foundation Award, NSF grant DBI-0960480 and NSF grant EF-1038697. GG was supported by the NSF through CAREER Grant No. DMR 09-55760.  DQ and AG would like to thank K.C. Huang for illuminating discussions. The authors would also like to acknowledge the hospitality of the Aspen Center for Physics (supported by NSF Grant No. PHY-1066293) where part of this work was done as well as the KITP workshop on ``Geometry, elasticity, fluctuations and order in 2D soft matter"  (NSF Grant No. PHY 11-25915), where this manuscript was completed.

\bibliography{ref_bj} 

\begin{thebibliography}{10}

\bibitem{cowin}
P.~Cowin and B.~Burke.
\newblock Cytoskeleton-membrane interactions.
\newblock {\em Curr.Op.Cell Biol.}, 8:56--65., 1996.

\bibitem{janmey}
R.~Lipowsky and E~Sackmann.
\newblock {\em Handbook of Biological Physics, 1: The Structure and dynamics of
  membranes. (Chapter 17)}.
\newblock Amsterdam: Elsevier Science, hoff a (ed) edition, 1995.

\bibitem{lamin}
N.~Stuurman, S.~Heins, and U.~Aebi.
\newblock Nuclear lamins: their structure, assembly and interactions.
\newblock {\em Journal of Structural Biology}, 122:42--46, 1998.

\bibitem{cont}
T.~Kamasaki, M.~Osumi, and I.~Mabuchi.
\newblock Three-dimensional arrangement of f-actin in the contractile ring of
  fission yeast.
\newblock {\em J. Cell Biol.}, 178:178:765171, 2007.

\bibitem{lloyd}
Clive Lloyd and Jordi Chan.
\newblock Microtubules and the shape of plants to come.
\newblock {\em Nature Reviews Molecular Cell Biology}, 5:13, 2004.

\bibitem{mreb}
F.~van~den Ent, C.M. Johnson, L.~Persons, P.~de~Boer, and J.~Lwe.
\newblock Bacterial actin mreb assembles in complex with cell shape protein
  rodz.
\newblock {\em EMBO J}, 29:10811790, 2010.

\bibitem{arkin}
S.S. Andrews and A.~P. Arkin.
\newblock A mechanical explanation for cytoskeletal rings and helices in
  bacteria.
\newblock {\em Biophys. J.}, 93:1872--1884, 2007.

\bibitem{mrebdir}
J.~Salje, F.~van~den Ent, P.~de~Boer, and J.~L{\"o}we.
\newblock Direct membrane binding by bacterial actin mreb.
\newblock {\em Mol. Cell}, 43:478177, 2011.

\bibitem{ftsz}
J.~Lutkenhaus.
\newblock Ftsz ring in bacterial cytokinesis.
\newblock {\em Molecular Microbiolog}, 9:403--9., 1993.

\bibitem{polyint}
G.~Fleer, M.~Cohen-Stuart, J.~Scheutjens, T.~Cosgrove, and B.~Vincent.
\newblock {\em Polymers at Interfaces}.
\newblock Chapman and Hall, London, 1993.

\bibitem{polyint2}
E.~Eisenriegler.
\newblock {\em Polymers at Interfaces}.
\newblock World Scientific, Singapore, 1993.

\bibitem{greg1}
Homin Shin and Gregory~M. Grason.
\newblock Structural reorganization of parallel actin bundles by crosslinking
  proteins: Incommensurate states of twist.
\newblock {\em Phys. Rev. E}, 82:051919, 2010.

\bibitem{greg2}
Homin Shin, Kristin.~R. Purdy, Drew. James, R.~Bartles, Gerard. C.~L. Wong, and
  Gregory~M. Grason.
\newblock Cooperativity and frustration in protein-mediated parallel actin
  bundles.
\newblock {\em Phys. Rev. Lett.}, 103:238102, 2009.

\bibitem{greg3}
Gregory~M. Grason.
\newblock Braided bundles and compact coils: The structure and thermodynamics
  of hexagonally packed chiral filament assemblies.
\newblock {\em Phys. Rev. E}, 79:041919, 2009.

\bibitem{greg4}
Gregory~M. Grason and Robijn~F. Bruinsma.
\newblock Chirality and equilibrium biopolymer bundles.
\newblock {\em Phys. Rev. Lett.}, 99:098101, 2007.

\bibitem{dqgreg}
David~A. Quint, Ajay Gopinathan, and Gregory~M. Grason.
\newblock Conformational collapse of surface-bound helical filaments.
\newblock {\em Soft Matter}, 8:9460, 2012.

\bibitem{sun_sm}
Nash~D. Rochman and Sean~X. Sun.
\newblock The twisted tauopathies: surface interactions of helically patterned
  filaments seen in alzheimer's disease and elsewhere.
\newblock {\em Soft Matter}, 12:779, 2016.

\bibitem{santoso}
Shuguang Zhang, Davide~M Marini, Wonmuk Hwang, and Steve Santoso.
\newblock Design of nanostructured biological materials through self-assembly
  of peptides and proteins.
\newblock {\em Current Opinion in Chemical Biology}, 6:865, 2002.

\bibitem{peptoid}
Ellen Robertson, Alessia Battigelli, Caroline Proulx, Ranjan Mannige, Thomas
  Haxton, Lisa Yun, Stephen Whitelam, and Ronald Zuckermann.
\newblock Design, synthesis, assembly and engineering of peptoid nanosheets.
\newblock {\em Accounts of Chemical Research}, 49:379, 2016.

\bibitem{amyloiddisease}
M.~Stefani.
\newblock Protein misfolding and aggregation: new examples in medicine and
  biology of the dark side of the protein world.
\newblock {\em Biochim. Biophys. Acta}, 1739:5, 2004.

\bibitem{amyloid_fret}
Galyna Gorbenko, Valeriya Trusova, Mykhailo Girych, Emi Adachi, Chiharu
  Mizuguchi, Kenichi Akaji, and Hiroyuki Saito.
\newblock Fret evidence for untwisting of amyloid fibrils on the surface of
  model membranes.
\newblock {\em Soft Matter}, 11:6223, 2015.

\bibitem{jen_pnas}
Jen Hsin, Ajay Gopinathan, and K.C. Huang.
\newblock Nucleotide-dependent conformations of ftsz dimers and force
  generation observed through molecular dynamics simulations.
\newblock {\em Proceedings of the National Academy of Sciences of the United
  States of America}, 109:9432, 2012.

\bibitem{tarazona_sm}
Pablo~Gonzalez de~Prado~Salas, Ines Horger, Fernando Martin-Garcia, Jesus
  Mendieta, Alvaro Alonso, Mario Encinar, Paulino Gomez-Puertas, Marisela
  Velez, and Pedro Tarazona.
\newblock Torsion and curvature of ftsz filaments.
\newblock {\em Soft Matter}, 10:1977, 2014.

\bibitem{alex_pnas}
Alexandre Colavin, Jen Hsin, and Kerwyn~Casey Huang.
\newblock Effects of polymerization and nucleotide identity on the
  conformational dynamics of the bacterial actin homolog mreb.
\newblock {\em Proceedings of the National Academy of Sciences of the United
  States of America}, 9:3585, 2014.

\bibitem{ftsz_jbc}
Victor~M. Hernandez-Rocamora, Belen Reija, Concepcion Garc{\'\i}a, Paolo
  Natale, Carlos Alfonso, Allen~P. Minton, Silvia Zorrilla, German Rivas, and
  Miguel Vicente.
\newblock Dynamic interaction of the escherichia coli cell division zipa and
  ftsz proteins evidenced in nanodiscs.
\newblock {\em Journal of Biological Chemistry}, 387:30097, 2012.

\bibitem{popp_jbc}
D.~Popp, M.~Iwasa, H.P. Erickson, A.~Narita, Y.~Maeda, and R.C. Robinson.
\newblock Suprastrcutures and dynamic properties of myobacterium tuberculsosis
  ftsz.
\newblock {\em J. Biol. Chem.}, 285:11281, 2010.

\bibitem{hamon_langmuir}
L.~Hamon, D.~Panda, P.~Savarin, V.~Joshi, J.~Bernhard, E.~Mucher, A.~Mechulam,
  P.~A. Curmi, and D.~Pastre.
\newblock Mica surface promotes the assembly of cytoskeletal proteins.
\newblock {\em Langmuir}, 25:3331--3335, 2009.

\bibitem{arumugam_angchem}
S.~Arumugam, G.~Chwastek, E.~Fischer-Friedrich, C.~Ehrig, I.~Monchand, and
  P.~Schwille.
\newblock Surface topology engineering of membranes for the mechanical
  investigation of the tubulin homologue ftsz.
\newblock {\em Angew. Chem., Int. Ed.}, 51:1185811862, 2012.

\bibitem{crescentin}
M.T. Cabeen, G.~Charbon, and C.~Jacobs-Wagner.
\newblock Bacterial cell curvature through mechanical control of cell growth.
\newblock {\em EMBO J}, 28:1208, 2009.

\bibitem{ariel_pnas}
Ariel Amir, Farinaz Babaeipour, Dustin~B McIntosh, David~R Nelson, and Suckjoon
  Jun.
\newblock Bending forces plastically deform growing bacterial cell walls.
\newblock {\em Proceedings of the National Academy of Sciences},
  111:5778--5783, 2014.

\bibitem{kc_gaussian}
Tristan~S Ursell, Jeffrey Nguyen, Russell~D Monds, Alexandre Colavin, Gabriel
  Billings, Nikolay Ouzounov, Zemer Gitai, Joshua~W Shaevitz, and Kerwyn~Casey
  Huang.
\newblock Rod-like bacterial shape is maintained by feedback between cell
  curvature and cytoskeletal localization.
\newblock {\em Proceedings of the National Academy of Sciences},
  11:E1025--E1034, 2014.

\bibitem{sun_bj}
Hongyuan Jiang, Fangwei Si, William Margolin, and Sean~X. Sun.
\newblock Mechanical control of bacterial cell shape.
\newblock {\em Biophysical Journal}, 101:327--335, 2011.

\end{thebibliography}
\bibliographystyle{unsrt} 

\end{document}